\documentstyle[12pt]{article}

\def\be{\begin{equation}}
\def\ee{\end{equation}}
\def\ba{\begin{array}{c}}
\def\ea{\end{array}}

\def\ben{$$}
\def\een{$$}
\begin{document}

\titlepage

 \begin{center}{\Large \bf
 A schematic model of scattering

  in
${\cal PT}-$symmetric Quantum Mechanics
 }\end{center}

\vspace{5mm}

 \begin{center}
Miloslav Znojil

\vspace{3mm}

 \'{U}stav jadern\'e fyziky AV \v{C}R,
250 68 \v{R}e\v{z}, Czech Republic

\vspace{5mm}

 {e-mail: znojil@ujf.cas.cz }

\end{center}

\vspace{5mm}

\section*{Abstract}

One-dimensional scattering problem admitting a complex, ${\cal
PT}-$symmetric short-range potential $V(x)$ is considered. Using a
Runge-Kutta-discretized version of Schr\"{o}dinger equation we
derive the formulae for the reflection and transmission
coefficients and emphasize that the only innovation emerges in
fact via a complexification of one of the potential-characterizing
parameters.

 \vspace{9mm}

\noindent
 PACS 03.65.Ge



\newpage

\section{Introduction}

Standard textbooks describe the stationary one-dimensional motion
of a quantum particle in a real potential well $V(x)$ by the
ordinary differential Schr\"{o}dinger equation
 \be
 \left [
 -\frac{d^2}{dx^2}+V(x) \right ]\,\psi(x)=E\,\psi(x)\,,
 \ \ \ \ \ \ \ x \in (-\infty,\infty)
 \label{SE}
 \ee
which may be considered and solved in the bound-state regime at
$E< V(\infty) \leq +\infty$ or in the scattering regime with, say,
$E=\kappa^2 > V(\infty) =0$. In this way one either employs the
boundary conditions $\psi(\pm \infty)=0$ and determines the
spectrum of bound states or, alternatively, switches to the
different boundary conditions, say,
 \be
 \psi(x) =
 \left \{
 \begin{array}{lr}
 A\,e^{i\kappa x}+B\,e^{-i\kappa x}\,,\ \ \ \ \
 & x \ll -1\,,\\
 C\,e^{i\kappa x}\,,\ \ \ \ \ \ \ \ \ \ \
 &x \gg 1\,.
 \ea
 \right .
 \label{scatbc}
 \ee
Under the conventional choice of $A=1$ the latter problem
specifies the reflection and transmission coefficients $B$ and
$C$, respectively \cite{Fluegge}.

The conventional approach to the quantum bound state problem has
recently been, fairly unexpectedly, generalized to many
unconventional and manifestly non-Hermitian Hamiltonians $H \neq
H^\dagger$ which are merely quasi-Hermitian, i.e., which are
Hermitian only in the sense of an identity $H^\dagger =
\Theta\,H\,\Theta^{-1}$ which contains a nontrivial ``metric"
operator $\Theta=\Theta^\dagger>0$ as introduced, e.g., in
ref.~\cite{Geyer}. The key ideas and sources of the latter new
development in Quantum Mechanics incorporate the so called ${\cal
PT}-$symmetry of the Hamiltonians and have been summarized in the
very fresh review by Carl Bender \cite{Carl}. This text may be
complemented by a sample \cite{proc} of the dedicated conference
proceedings.

In this context we intend to pay attention to a very simple ${\cal
PT}-$symmetric {\em scattering} model where
 \ben
 V(x) = Z(x) + i\,Y(x)\,,
 \ \ \ \ \
 Z(-x)=Z(x)= {\rm real}\,,\ \ \ \ \ Y(-x)=-Y(x)= {\rm real}\,
 \een
and where the ordinary differential equation (\ref{SE}) is
replaced by its Runge-Kutta-discretized, difference-equation
representation
 \be
 -\frac{\psi(x_{k-1})-2\,\psi(x_k)+\psi(x_{k+1})}{h^2}+V(x_k)\,
 \psi(x_k)
 =E\,\psi(x_k)\,
 \label{SEdis}
 \ee
with
 \ben
  x_k=k\,h\,,\ \ \ \ \ h > 0\,,
 \ \ \ \ \ k = 0, \pm 1, \ldots
 \een
as employed, in the context of the bound-state problem, in refs.
\cite{RKBS}.

\section{Runge-Kutta scattering}

Once we assume, for the sake of simplicity, that the potential in
eq.~(\ref{SEdis}) vanishes beyond certain distance from the
origin,
 \ben
 V(x_{\pm j})=0\, \ \ \ \ j = M, M+1, \ldots\,,
 \een
we may abbreviate $\psi_k=\psi(x_k)$, $V_k={h}^2 V(x_k)$ and
$2\,\cos \varphi=2-h^2 E$ in eq.~(\ref{SEdis}),
 \be
 -\psi_{k-1}+\left (2\,\cos \varphi+V_k\right )\,\psi_k
 -\psi_{k+1}=0\,.
 \label{SEdisc}
 \ee
In the region of $|k| \geq M$ with vanishing potential $V_k=0$ the
two independent solutions of our difference Schr\"{o}dinger
eq.~(\ref{SEdisc}) are easily found, via a suitable ansatz, as
elementary functions of the new ``energy" variable $\varphi$,
 \ben
 \psi_k=const\,\cdot\,\varrho^k\ \ \Longrightarrow \ \ \varrho=\varrho_{\pm}=
 \exp(\pm i\,\varphi)\,.
 \een
This enables us to replace the standard boundary conditions
(\ref{scatbc}) by their discrete scattering version
 \be
 \psi(x_m) =
 \left \{
 \begin{array}{lr}
 A\,e^{i\,m\,\varphi}+B\,e^{-i\,m\,\varphi}\,,\ \ \ \ \
 & m \leq -(M-1)\,,\\
 C\,e^{i\,m\,\varphi}\,,\ \ \ \ \ \ \ \ \ \ \
 &m \geq M-1\,
 \ea
 \right .
 \label{discatbc}
 \ee
with a conventional choice of $A=1$.

Two comments may be added here. Firstly, one notices that the
condition of the reality of the new energy variable $\varphi$
imposes the constraint upon the original energy itself, $-2\leq
2-h^2 E \leq 2$, i.e., $E \in (0, 4/h^2)$. At any finite choice of
the lattice step $h>0$ this inequality is intuitively reminiscent
of the spectra in relativistic quantum systems. Via an explicit
display of the higher ${\cal O}(h^4)$ corrections in
eq.~(\ref{SEdis}), this connection has been given a more
quantitative interpretation in ref.~\cite{jaotom}.

The second eligible way of dealing with the uncertainty
represented by the ${\cal O}(h^4)$ discrepancy between the
difference- and differential-operator representation of the
Schr\"{o}dinger's kinetic energy is more standard and lies in its
disappearance in the limit $h \to 0$. This is a purely numerical
recipe known as the Runge-Kutta method \cite{Acton}. In the
present context of scattering one has to keep in mind that the two
``small" parameters $h$ and $1/M$ may and, in order to achieve the
quickest convergence, should be chosen and varied independently.

\section{The matching method of solution}

\subsection{The simplest model of the scattering with $M=1$ }

Once we are given the boundary conditions (\ref{discatbc}) the
process of the construction of the solutions is straightforward.
Let us first illustrate its key technical ingredients on the model
with the first nontrivial choice of the cutoff $M=1$. In this case
our difference Schr\"{o}dinger eq.~(\ref{SEdisc}) degenerates to
the mere three nontrivial relations,
 \be
 \ba
 -\psi_{-2}+2\,\cos \varphi\,\psi_{-1}
 -\psi_{0}^{(-)}=0\,\\
 -\psi_{-1}+\left (2\,\cos \varphi+Z_0\right )\,\psi_0
 -\psi_{1}=0\,\\
 -\psi_{0}^{(+)}+2\,\cos \varphi\,\psi_{1}
 -\psi_{2}=0\,
 \ea
 \ee
where we may insert, from eq.~(\ref{discatbc}),
 \be
 \psi_{-1} =
  e^{-i\,\varphi}+B\,e^{i\,\varphi}\,,\ \ \ \ \
 \psi_{0}^{(-)} =
  1+B\,,\ \ \ \ \ \
 \psi_{0}^{(+)} =
 C\,,\ \ \ \ \ \ \
 \psi_{1} =
 C\,e^{i\,\varphi}\,
 \ee
and where we have to demand, subsequently,
 \be
 \ba
 \psi_{0}^{(-)}=1+B =\psi_{0}^{(+)}=C =\psi_{0}\,,\\
 - e^{-i\,\varphi}-B\,e^{i\,\varphi}
 +\left (2\,\cos \varphi+Z_0\right )\,C
 -C\,e^{i\,\varphi}=0
 \,.
 \ea
 \ee
Thus, at an arbitrary ``energy" $\varphi$ one identifies $B = C-1$
and gets the solution
 \ben
 C=\frac{2i\sin\varphi }{2i\sin\varphi -Z_0}\,,\ \ \ \ \ \
 B=\frac{Z_0}{2i\sin\varphi -Z_0}\,.
 \een
Of course, as long as we deal just with the real ``interaction
term" $Z_0$, our $M=1$ toy problem remains Hermitian since no
${\cal PT}-$symmetry has entered the scene yet.

\subsection{${\cal PT}-$symmetry and the scattering at $M=2$ }

In the next, $M=2$ version of our model we have to insert the four
known quantities
 \ben
 \psi_{-2} =
  e^{-2\,i\,\varphi}+B\,e^{2\,i\,\varphi}\,,\ \ \ \ \
 \psi_{-1} =
  e^{-i\,\varphi}+B\,e^{i\,\varphi}\,,\ \ \ \ \
 \psi_{1} =
 C\,e^{i\,\varphi}\,, \ \ \ \ \
 \psi_{2} =
 C\,e^{2\,i\,\varphi}\,
 \een
in the triplet of relations
 \be
 \ba
 -\psi_{-2}+\left (2\,\cos \varphi+Z_{-1}-i\,Y_{-1}
 \right )\,\psi_{-1}
 -\psi_{0}^{(-)}=0\,\\
 -\psi_{-1}+\left (2\,\cos \varphi+Z_0\right )\,\psi_0
 -\psi_{1}=0\,\\
 -\psi_{0}^{(+)}+\left (2\,\cos \varphi+Z_{-1}+i\,Y_{-1}
 \right )\,\psi_{1}
 -\psi_{2}=0\,
 \ea
 \label{susu}
 \ee
where the three symbols $\psi_0$, $\psi_{0}^{(-)}$ and
$\psi_{0}^{(+)}$ defined by these respective equations should
represent the same quantity and must be equal to each other,
therefore. Having this in mind we introduce $\xi_0^{(-)}= 1+B$ and
$\xi_0^{(+)}=C$ and decompose
 \ben
 \psi_{0}^{(-)}=\xi_0^{(-)}+\chi_{0}^{(-)}\,,\ \ \ \ \
 \psi_{0}^{(+)}=\xi_0^{(+)}+\chi_{0}^{(+)}\,.
 \een
This enables us eliminate
 \ben
 \chi_{0}^{(-)}=V_{-1}\,\psi_{-1}\,,
 \ \ \ \ \ \
 \chi_{0}^{(+)}=V_{1}\,\psi_{1}\,
 \een
and eq.~(\ref{susu}) becomes reduced to the pair of conditions,
 \be
 \ba
 1+B+V_{-1}\,\psi_{-1}=C+V_{1}\,\psi_{1}=\psi_0\,,\\
 -\psi_{-1}+\left (2\,\cos \varphi+Z_0\right )\,\psi_0
 -\psi_{1}=0\,
 \ea
 \ee
They lead to the two-dimensional linear algebraic problem which
defines the reflection and transmission coefficients $B$ and $C$
at any input energy $\varphi$. The same conclusion applies to all
the models with the larger $M$.

\section{The matrix-inversion method of solution}


Let us now re-write our difference Schr\"{o}dinger
eq.~(\ref{SEdisc}) as a doubly infinite system of linear algebraic
equations
 \be
 \left (
 \begin{array}{ccccc}
 \ddots&\ddots&\ddots &\vdots&\\
 \ddots&S_{-1}&-1& 0&\ldots \\
 \ddots&-1&S_{0}&-1&\ddots\\
  \ldots&0&-1&S_1&\ddots\\
 & \vdots& \ddots&\ddots&\ddots
 \ea
 \right )
 \,\left (
 \ba
 \vdots\\
 \psi_{-1}\\
 \psi_0\\
 \psi_1\\
 \vdots
 \ea
 =\right )=0\,,
 \label{bigmat}
 \ee
where
 \be
 S_k\ (\ \equiv\  S_{-k}^*)\ =\left \{
 \begin{array}{lr}
 2\,\cos \varphi +Z_k+{\rm i}\,Y_k\,{\rm sign}\, k\,,\ \ &
 |k|<M\,,\\
 2\,\cos \varphi\,,
 &|k| \geq M\,
 \ea
 \right .
 \ee
and where the majority of the elements of the ``eigenvector" are
prescribed, in advance, by the boundary conditions
(\ref{discatbc}). Once we denote all of them by a different
symbol,
 \be
 \psi(x_m) =
 \left \{
 \begin{array}{rr}
 A\,e^{i\,m\,\varphi}+B\,e^{-i\,m\,\varphi}\ \equiv\ \xi_m^{(-)}\,,
 & \ \ \ \ \ \  m \leq -(M-1)\,,\\
 C\,e^{i\,m\,\varphi}\ \equiv\ \xi_m^{(+)}\,,
 &m \geq M-1\,,
 \ea
 \right .
 \label{discatbcdr}
 \ee
we may reduce eq.~(\ref{bigmat}) to a finite-dimensional and
tridiagonal non-square-matrix problem
 \be
 \left (
 \begin{array}{ccccccccc}
 -1&S_{(M-1)}^*&-1&&&&&&\\
 &\ddots&\ddots&\ddots& &&&&\\
 &&-1&S_{1}^*&-1& &&&\\
 &&&-1&S_{0}&-1&&&\\
 && &&-1&S_1&-1&\ddots &  \\
 &&& & \ddots&\ddots&\ddots&\ddots&\\
 &&&&&&-1&S_{(M-1)}&-1
 \ea
 \right )
 \,\left (
 \ba
  \xi_{-M}^{(-)}
 \\
  \xi_{-(M-1)}^{(-)}
 \\
 \psi_{-(M-2)}\\
 \vdots\\
 \psi_{M-2}\\
  \xi_{M-1}^{(+)}
 \\
  \xi_M^{(+)}
  \ea
 =\right )=0\,
 \label{midmat}
 \ee
or, better, to a non-homogeneous system of $2M-1$ equations
 \be
 \mbox{\rm \large \bf T }\cdot
 \,\left (
 \ba
  \xi_{-(M-1)}^{(-)}
 \\
 \psi_{-(M-2)}\\
 \vdots\\
 \psi_{M-2}\\
  \xi_{M-1}^{(+)}
  \ea
 \right )=
\left (
 \ba
  \xi_{-M}^{(-)}\\
 0
 \\
 \vdots\\
 0
 \\
  \xi_M^{(+)}
  \ea
 \right )\,
 \label{smallmat}
 \ee
where the $(2M-1)-$dimensional square-matrix of the system can be
partitioned as follows,
 \be
 \mbox{\rm \large \bf T }\ \mbox{\rm \Large  \  = } \  \left (
 \begin{array}{c|ccccc|c}
 S_{(M-1)}^*&-1&&&&&\\
 \hline
 -1&S_{(M-2)}^*&-1& &&&\\
 &-1&\ddots&\ddots& &&\\
 &&\ddots&S_{0}&\ddots&&\\
 & &&\ddots&\ddots&-1&   \\
 && &&-1&S_{(M-2)}&-1\\
 \hline
 &&&&&-1&S_{(M-1)}
 \ea
 \right )
 \,.
  \label{defmat}
 \ee
Whenever this matrix proves non-singular, it may assigned the
inverse matrix {\bf R}={\bf T}$^{-1}$, the knowledge of which
enables us to re-write eq.~(\ref{smallmat}), in the same
partitioning, as follows,
 \be
 \left (
 \ba
  \xi_{-(M-1)}^{(-)}
 \\
 \hline
 \vec{\Psi}\\
 \hline
  \xi_{M-1}^{(+)}
  \ea
 \right )=
  \mbox{\rm \large \bf R }\cdot
\left (
 \ba
  \xi_{-M}^{(-)}\\
 \vec{0}
 \\
  \xi_M^{(+)}
  \ea
 \right )\,,\ \ \ \ \ \
 \vec{\Psi}= \left (
 \ba
 \psi_{-(M-2)}\\
 \vdots\\
 \psi_{M-2}
  \ea
 \right )\,.
 \label{invmat}
 \ee
In the next step we deduce that the matrix {\bf R} has the
following partitioned form
 \ben
 \mbox{\rm \large \bf R }\ \mbox{\rm \Large  \  = } \  \left (
 \begin{array}{c|c|c}
 \alpha^*&\vec{t}^T&\beta\\
 \hline
 \vec{u}&{\rm \bf Q}&\vec{v}\\
  \hline
 \beta&\vec{w}^T&\alpha
 \ea
 \right )
 \,.
 \een
We may summarize that in the light of the overall partitioned
structure of eq.~(\ref{invmat}), the knowledge of the
$(2M-3)-$dimensional submatrix {\rm \bf Q} as well as of the two
$(2M-3)-$dimensional row vectors $\vec{t}^T$ and $\vec{w}^T$
(where $^T$ denotes transposition) is entirely redundant.
Moreover, the knowledge of the other two column vectors $\vec{u}$
and $\vec{v}$ only helps us to eliminate the ``wavefunction"
components $\psi_{-(M-2)}$, $\psi_{-(M-3)}$, $\ldots,$
$\psi_{M-3}$, $\psi_{M-2}$. In this sense, equation
(\ref{smallmat}) degenerates to the mere two scalar relations
 \be
 \ba
 \xi_{-(M-1)}^{(-)}-\alpha^*\,
  \xi_{-(M)}^{(-)}
 -\beta\,
   \xi_{M}^{(+)}=0\,,
   \\
    \xi_{M-1}^{(+)}
 -\beta\,
   \xi_{-M}^{(-)}
 -\alpha\,
    \xi_{M}^{(+)}=0\,.
    \ea
    \label{fineq}
 \ee
Once we insert the explicit definitions from
eq.~(\ref{discatbcdr}) we get the final pair of linear equations
 \be
 \ba
 e^{-i\,(M-1)\,\varphi}+B\,e^{i\,(M-1)\,\varphi}
 -\alpha^*\,
 \left (e^{-i\,M\,\varphi}+B\,e^{i\,M\,\varphi}\right )
 -C\,\beta\,
 e^{i\,M\,\varphi}
   =0\,,
   \\
 C\,e^{i\,(M-1)\,\varphi}
  -\beta\,
 \left (e^{-i\,M\,\varphi}+B\,e^{i\,M\,\varphi}\right )
 -C\,\alpha\,
 e^{i\,M\,\varphi}
    =0\,
    \ea
    \label{finexp}
 \ee
which are solved by the elimination of
 \be
 B=-e^{-2iM\varphi}+\frac{C}{\beta}\,\left (e^{-i\varphi}-\alpha
 \right )
 \label{becko}
 \ee
and, subsequently, of
 \be
 C=
  \frac{2i\beta e^{-2i M\varphi}\sin \varphi}
  {\beta^2-(e^{-i\varphi}-\alpha^*)\,(e^{-i\varphi}-\alpha)}\,.
  \label{cecko}
  \ee
This is our present main result.

\section{Coefficients $\alpha$ and $\beta$ \label{last} }

Our final scattering-determining formulae (\ref{becko}) and
(\ref{cecko}) indicate that the complex coefficient $\alpha$ and
the real coefficient $\beta$ carry all the ``dynamical input"
information. At any given energy parameter $\varphi$ these matrix
elements are, by construction, rational functions of our $2M-1$
real coupling constants $Z_0, Z_1, \ldots, Z_{M-1}$ and $Y_1,
\ldots,Y_{M-1}$. In particular, $\beta$ is equal to $1/\det {\rm
\bf T} $ and $\alpha$ has the same denominator of course. An
explicit algebraic determination of the determinant $\det {\rm \bf
T} $ and of the numerator (say, $\gamma$) of $\alpha$ is less
easy. Let us illustrate this assertion on a few examples.

\subsection{$M=2$ once more}

 \ben
 \det {\rm \bf T}=
 {\it {Z}_0}\,{{\it {Z}_1}}^{2}-2\,{\it {Z}_1}+{{\it Y_1}}^{2}{\it {Z}_0}
 \een

${\rm Re}\,\gamma={\it {Z}_0}\,{\it {Z}_1}-1$

${\rm Im}\,\gamma=-{\it {Z}_0}\,{\it Y_1}$

\subsection{$M=3$}

 \ben
 \det {\rm \bf T}=
{\it {Z}_0}\,{{\it {Z}_1}}^{2}{{\it {Z}_2}}^{2}-2\,{\it
{Z}_0}\,{\it {Z}_1}\,{\it {Z}_2} -2\,{\it {Z}_1}\,{{\it
{Z}_2}}^{2}+
 \een
  \ben
  +{{\it Y_1}}^{2}{\it {Z}_0}\,{{\it
{Z}_2}}^{2}+2 \,{\it {Z}_2}+{{\it Y_2}}^{2}{\it {Z}_0}\,{{\it
{Z}_1}}^{2}-2\,{{\it Y_2}}^{2}{ \it {Z}_1}+{{\it Y_2}}^{2}{{\it
Y_1}}^{2}{\it {Z}_0}+2\,{\it {Z}_0}\,{\it Y_1}\,{ \it Y_2}+{\it
{Z}_0}
 \een

${\rm Re}\,\gamma={\it {Z}_0}\,{{\it {Z}_1}}^{2}{\it
{Z}_2}-2\,{\it {Z}_1}\,{\it {Z}_2}+{{\it Y_1}}^{2}{ \it
{Z}_0}\,{\it {Z}_2}-{\it {Z}_1}\,{\it {Z}_0}+1$

${\rm Im}\,\gamma=-{\it {Z}_0}\,{{\it {Z}_1}}^{2}{\it Y_2}+2\,{\it
{Z}_1}\,{\it Y_2}-{{\it Y_1}}^{2} {\it {Z}_0}\,{\it Y_2}-{\it
Y_1}\,{\it {Z}_0}$

\subsection{$M=4$}

The growth of complexity of the formulae occurs already at the
dimension as low as $M=4$. The determinant $\det {\rm \bf T} $ and
the real and imaginary parts of $\gamma$  are them represented by
the sums of 15 and 14 and 32 products of couplings, respectively.

A simplification is only encountered in the weak coupling regime
where one finds just two terms in the determinant which are linear
in the couplings,
 \ben
 \det {\rm \bf T}=-2\,{\it {Z}_3}-2\,{\it {Z}_1}+ \ldots
  \een
being followed by the 10 triple-product terms,
  \ben
 \ldots + {{\it Y_1}}^{2}{\it {Z}_0
}+4\,{\it {Z}_1}\,{\it {Z}_2}\,{\it {Z}_3}-4\,{\it {Z}_1}\,{\it
Y_2}\,{\it Y_3}+{{ \it {Z}_1}}^{2}{\it {Z}_0}+2\,{{\it
Y_3}}^{2}{\it {Z}_2}+2\,{\it Y_1}\,{\it {Z}_0}\, {\it Y_3}+
 \een
 \ben
 +2\,{\it
{Z}_1}\,{\it {Z}_0}\,{\it {Z}_3}+{{\it {Z}_3}}^{2}{\it
{Z}_0}+{{\it Y_3}}^{2}{\it {Z}_0}+2\,{{\it {Z}_3}}^{2}{\it {Z}_2}
+\ldots
 \een
etc. Similarly, we may decompose, in the even-number products,
 \ben
 {\rm Re}\,\gamma=-1+2\,{\it {Z}_2}\,{\it {Z}_3}+{\it
{Z}_0}\,{\it {Z}_1}+{\it {Z}_0}\,{\it {Z}_3} +2\,{\it {Z}_2}\,{\it
{Z}_1}+ \ldots
 \een
and continue
 \ben
 \ldots +-2\,{\it {Z}_0}\,{\it
{Z}_1}\,{ \it {Z}_2}\,{\it {Z}_3}+2\,{\it {Z}_0}\,{\it Y_1}\,{\it
Y_2}\,{\it {Z}_3}-
 \een
 \ben
 -{\it {Z}_2}\, {{\it Y_1}}^{2}{\it
{Z}_0}-2\,{{\it {Z}_2}}^{2}{\it {Z}_1}\,{\it {Z}_3}-{\it {Z}_2}\,{
{\it {Z}_1}}^{2}{\it {Z}_0}-2\,{{\it Y_2}}^{2}{\it {Z}_1}\,{\it
{Z}_3} +\ldots
 \een
etc, plus
 \ben
 {\rm Im}\,\gamma=-2\,{\it {Z}_2}\,{\it Y_3}+2\,{\it
Y_2}\,{\it {Z}_1}-{\it {Z}_0}\,{\it Y_1 }-{\it {Z}_0}\,{\it Y_3} -
\ldots
 \een
with a continuation
 \ben
 \ldots -{\it Y_2}\,{{\it Y_1}}^{2}{ \it {Z}_0}-{\it Y_2}\,{{\it
{Z}_1}}^{2}{\it {Z}_0}+
 \een
 \ben
 +2\,{{\it Y_2}}^{2}{\it {Z}_1}\,{ \it
Y_3}+2\,{\it {Z}_0}\,{\it {Z}_1}\,{\it {Z}_2}\,{\it Y_3}+2\,{{\it
{Z}_2}}^{2}{ \it {Z}_1}\,{\it Y_3}-2\,{\it {Z}_0}\,{\it Y_1}\,{\it
Y_2}\,{\it Y_3}    +\ldots
 \een
etc. Symbolic manipulations on a computer should be employed at
all the higher dimensions $M\geq 4$ in general.

\section{Discussion}

The main inspiration of the activities and attention paid to the
${\cal PT}-$symmetry originates from the pioneering 1998 letter by
Bender and Boettcher \cite{BB} where the operator ${\cal P}$ meant
parity and where the (antilinear) ${\cal T}$ represented time
reversal. Its authors argued that the {\em complex} model $V(x) =
x^2\,(ix)^\delta$ seems to possess the {\em purely real}
bound-state spectrum at all the exponents $\delta\geq 0$. After a
rigorous mathematical proof of this conjecture by Dorey, Duncan,
Tateo and Shin \cite{DDT} and after the (crucial) clarification of
the existence of a nontrivial, ``physical" Hilbert space ${\cal
H}$ where the Hamiltonian remains self-adjoint
\cite{Geyer,pseudo,ali,BBJ}, the bound-state version of
eq.~(\ref{SE}) may be considered more or less well understood,
especially after it has been clarified that the physics-inspired
concept of ${\cal PT}-$symmetry of a Hamiltonian $H$ should in
fact be understood, in the language of mathematics, as a ${\cal
P}-$pseudo-Hermiticity of $H$ specified by the property
$H^\dagger={\cal P}\,H\,{\cal P}^{-1}$
\cite{pseudo,ali,solombrino}.

In the spirit of the latter generalization, current literature
abounds in the studies of the potentials which are analytically
continued \cite{BBjmp}, singular and multisheeted \cite{tobog},
multidimensional \cite{dimen}, manybody \cite{ag3}, relativistic
\cite{KG}, supersymmetric \cite{susy} and channel-coupling
\cite{cc}. Among all these developments, a comparatively small
number of papers has been devoted to the problem of the
scattering. For a sample one might recollect the key reviews
\cite{Cannatab} and various Kleefeld's conceptual conjectures
\cite{Kleefeld} as well as a very explicit study of the scattering
by the separable ${\cal PT}-$symmetric potentials of rank one
\cite{Cannata} or by the rectangular or reflectionless barriers
\cite{Ahmed}, or the motion considered along the so called
tobogganic (i.e., complex {\em and} topologically nontrivial)
integration contours \cite{tobog2}. In this context our present
difference-equation-based study may be understood just as another
attempt to fill the gap.

Technically we felt inspired by our old Runge-Kutta-type
discretization of the ${\cal PT}-$symmetric Schr\"{o}dinger
equations \cite{RKBS} as well as by our recent chain-model
approximations of bound states in a finite-dimensional Hilbert
space \cite{findim}. In a certain unification of these two
approaches we succeeded here in showing that there exists a close
formal parallelism between the description of the
(one-dimensional, Runge-Kutta-approximated) scattering by a real
(i.e., Hermitian) potentials and by their complex, ${\cal
PT}-$symmetric generalizations. We showed that in both these
contexts, the definition of the transmission and reflection
coefficients has the same form [cf., e.g., eq.~(\ref{cecko})],
with all the differences represented by the differences in the
form of the ``dynamical input" information. It has been shown to
be encoded,  in {\em both} the Hermitian and non-Hermitian cases,
in the two functions $\alpha$ and $\beta$ of the lattice
potentials, with the vanishing or non-vanishing coefficients
$Y_k$, respectively (cf. a few samples of the concrete form of
$\alpha$ and $\beta$ in section \ref{last}).

On the level of physics we would like to emphasize that one of the
main distinguishing features of the scattering problem  in ${\cal
PT}-$symmetric quantum mechanics lies in the manifest asymmetry
between the ``in" and ``out" states \cite{Kleefeld}. In its
present solvable exemplification we showed that such an asymmetry
is merely formal and that the problem remains tractable by the
standard, non-matching and non-recurrent techniques of linear
algebra. A key to the success proved to lie in the partitioning of
the Schr\"{o}{dinger equation which enabled us to separate its
essential and inessential components and to reduce the
construction of the amplitudes to the mere two-dimensional matrix
inversion [cf. eq.~(\ref{fineq})] where all the dynamical input is
represented by the four corners of the inverse matrix {\bf R} =
{\bf T}$^{-1}$  [cf. the definition (\ref{defmat})].

We believe that the merits of the present discrete model were not
exhausted by its present short analysis and that its further study
might throw new light, e.g., on the non-Hermitian versions of the
inverse problem of scattering. \vspace{5mm}

\section*{Acknowledgement}

Work supported by the M\v{S}MT ``Doppler Institute" project Nr.
LC06002,  by the Institutional Research Plan AV0Z10480505 and by
the GA\v{C}R grant Nr. 202/07/1307.

\end{document}